\documentclass[final,5p,times,twocolumn]{elsarticle}
\usepackage{amsbsy}
\usepackage{amssymb}
\usepackage{amsmath}
\newcommand{\beq}{\begin{equation}}
\newcommand{\eeq}{\end{equation}}
\newcommand{\be}{\begin{eqnarray}}
\newcommand{\ee}{\end{eqnarray}}

\journal{Physics Letters B}

\begin{document}
\begin{frontmatter}
\title{On the production of heavy axion-like particles in the accretion disks of gamma-ray bursts}
\author[label1]{Mat\'{\i}as M. Reynoso}

\address[label1]{IFIMAR (CONICET-UNMdP) and Departamento de F\'{\i}sica, Facultad de Ciencias Exactas y Naturales, Universidad Nacional de Mar del Plata, Funes 3350, (7600) Mar del Plata, Argentina}
 \ead{mreynoso@mdp.edu.ar}
\begin{abstract}
{Heavy axion-like particles have been introduced in several scenarios beyond the Standard Model and their production {should be possible} in some astrophysical systems. In this {study}, we re-examine the possibility that {this type of particle} can be generated in the accretion disks of gamma-ray bursts (GRB), which are the most powerful events in the universe. If the produced axions decay into photons or $e^+e^-$ pairs at the correct distances, a fireball is generated. We calculate the structure {of} transient accretion disks in GRBs (density, temperature and thickness profiles) {considering the effect} of heavy axion emission as well as the rest of the relevant standard cooling processes. This allows us to obtain the values of the coupling constant $g_{aN}$ {such that the axions do not} become trapped, and we also compute {the heavy axion luminosity emitted} from the entire disk. We {show} that for the couplings within the ranges found, the mechanism for powering GRBs based on heavy axion production and decay {is} an alternative to the standard picture based {on} magnetohydrodynamic processes and neutrino-antineutrino annihilation. {Alternatively, the mechanism fails} if heavy axions are produced in the disk but their decay takes place further away. Still, the decay products (gamma rays or electrons and positrons) should leave observable signatures, which are not observed for different ranges of values of the coupling constants, depending on the mass of the heavy axion.
}
\end{abstract}
\begin{keyword}
new physics \sep axions \sep gamma-ray bursts

\end{keyword}
\end{frontmatter}

\section{Introduction}\label{sec:intro}

Gamma-ray bursts (GRBs) are among the most powerful events in the universe since the Big-Bang, emitting radiation at a rate  $L_\gamma\sim 10^{51-53}{\rm erg\ s^{-1}}$, {which} is observed on Earth as flashes of gamma rays with energies $\sim (10 \,  {\rm keV} - 50 {\rm MeV})$ {and durations ranging} from a fraction to hundreds of seconds (for a review, see  \citep{meszaros2006}). {GRBs that} last less than two seconds are considered short GRBs and they are thought to be produced by the merging of compact objects in a
binary system \cite{pacynski1986}. {Longer GRBs are associated with} the collapse of
a massive star into a black hole \citep{woosley1993}.  It is {generally} accepted that an initial fireball of $e^+e^-$, gamma rays, and baryons is generated close to the black hole, {which then expands to reach} ultrarelativistic velocities \citep{cavallorees1978}. Hence, the observed radiation is {considered} to correspond to synchrotron and/or inverse Compton (IC) emission of electrons that
have been accelerated to high energies in shocks created by relativistic plasma
shells that collide \citep{reesmeszaros1994,piran1999}. GRBs {{can also} be detected by observing their} lower energy afterglow emission, {which} occurs from hours to days after the initial detection {and it} can last for months in some cases. This has allowed to measure the corresponding redshift $z$, {thereby confirming} the extragalactic origin of GRBs (e.g. {see} \cite{afterglowdetection1997,afterglowdetection1998}). 

As for the central engine, the black hole is {assumed} to be surrounded by a transient, hot, and dense accretion disk{, which is considered to be cooled via advection} and neutrino emission \citep{pwf1999,kohri2002,dimatteo2002,kohri2005,chen2007}. The high densities and temperatures in this disk {as well as} in the generated fireball have motivated the study of different particle physics beyond the Standard Model \cite{loeb1993,bertolami1998,demir1999,viaaxion2000,mirror2004,tu2015}. {In this study}, we further explore the possibility that heavy axion-like particles can be emitted from such accretion disks \citep{viaaxion2000, mirror2004,nucleonbrem2005}.

The existence of the axion was proposed to solve the strong CP problem, which {(for example) is reflected by} the fact that the electric dipole moment of the neutron is unnaturally small \cite{lowenergyfrontier2010}. As a possible solution to this problem, the \textit{standard axion} arises as a {pseudo-Nambu--Goldstone} boson that spontaneously breaks the {Peccei--Quinn} symmetry ($U_{\rm PQ}$) at a scale $f_a$ \cite{pq1977}. This cancels out the CP-violating term of the QCD Lagrangian and the axion acquires a mass given by
\beq
m_a\simeq 6\times 10^{-4}{\rm eV}\left(\frac{10^{10}{\rm GeV}}{f_a}\right)\label{axionrel}.
\eeq
Heavy axion-like particles\footnote{{Throughout} the text we refer to ``heavy axion-like particles" just as ``heavy axions".} arise if the above condition for the mass is relaxed, and although the strong CP problem {may} not be solved, it is an interesting possibility that such heavy axions may indeed exist, as has been proposed in the context of several theoretical scenarios beyond the Standard Model \citep{grandunification1997,extradim1999,strongcp2000}. {The Lagrangian terms describing the heavy axion interaction with matter are given by} \citep{viaaxion2000}
\beq
\mathcal{L}_a= -\frac{1}{2}m_a^2a^2-ig_{ai}a\bar{\psi}_i\gamma_5 {\psi}_i-\frac{g_{a\gamma}}{4}aF_{\mu\nu}\tilde{F}^{\mu\nu},
\eeq 
where 
$g_{ai}$ are the coupling constants for interactions with fermions $\psi_i$, $(i=e,N)$ with $N=(p,n)$, and
$g_{a\gamma}$ 
 {parameterizes} the coupling to photons. In particular, the most stringent bounds {on} the coupling constants $g_{aN}$ and $g_{a\gamma}$ refer to the standard (light) axion, for which both constants are related and relation (\ref{axionrel}) holds \cite{lowenergyfrontier2010,khlopov1978,khlopov1991,khlopov1992}. 
For instance, bounds on $g_{a\gamma}$ were obtained {based on} the duration of {SN 1987A} \cite{masso1995}, i.e. $g_{a\gamma}\lesssim 10^{-6}\,{\rm GeV^{-1}}$ for $m_a\sim 1-10\,{\rm MeV}$, and {a} more recent analysis \cite{jaeckel2017} {concluded} that $g_{a\gamma}\lesssim 10^{-14}\,{\rm GeV^{-1}}$ for the same range of masses. 
{In addition}, studying the production of heavy axions inside a supernova core leads to constraints on $g_{aN}$, as discussed {by} \citep{newconstraints2011} in the case {where the axions produced} can escape freely outside\footnote{{In particular}, the results {obtained by} \cite{nucleonbrem2005} imply that the mean free path for axions of mass $\sim 1$ MeV {will} be larger than the core radius $R_{\rm SN}=10$ km if $g_{aN}<10^{-8}$ for a core density $\rho_{\rm SN}\approx 10^{14}{\rm g \ cm^{-3}}$ and temperature $T_{\rm SN}\approx 30{\,\rm MeV}$.}. Further bounds on the coupling to photons $g_{a\gamma}$ can be derived from {cosmological} considerations, i.e., heavy axions created in the early universe would decay, {thereby affecting} the cosmic microwave background (CMB) and the extragalactic background light (EBL), and such decays could also dilute the neutrino density or create a diffuse photon background \cite{cadamuro2012,millea2015}.

 {In this study}, as in \cite{viaaxion2000}, we {use} a phenomenological approach and {maintain all of the} coupling constants as independent.  
{We {consider} heavy axions of mass $m_a=0.1-10$ MeV which can couple to photons, electrons, and nucleons, adopting for the latter values in the range $g_{aN}\sim 10^{-7}-10^{-5}$, {and we make} no distinction between neutrons and protons. We {give} two reasons why this range {is suitable for consideration}. {First}, there is an allowed window for $g_{aN}\sim (1 - 3) \times 10^{-6}$ \cite{hewett2012,pdg2016}, and {second}, the existing bound excluding the range $(3\times 10^{-9}\lesssim g_{aN}\lesssim 10^{-6})$ is derived from the observed duration of {SN 1987A}, {although based on data with} poor statistics, and without a complete understanding of the dynamics of the explosion (e.g., {see }\cite{cadamuro2012}). Hence, it may be interesting to explore further astrophysical phenomena in which such values of $g_{aN}$ can also be tested. Under {the assumptions stated above}, the leading process for heavy axion production is {nucleon--nucleon} bremsstrahlung \citep{raffeltlab1996}, and considering non-zero couplings to leptons and/or photons {allows} the possibility of decays to these particles, which may have observable signatures under appropriate conditions. Studying these effects  }  
in the context of {highly plausible scenarios that are} considered appropriate for the generation of GRBs, 
our work provides complementary results {to the existing constraints, {which} can be expressed as} 
restrictions on the $g_{a\gamma}-g_{ae}$ and $g_{a\gamma}-m_{a}$ planes for different values of $g_{aN}$. 
        
{In particular}, we focus on the production of heavy axions in the accretion disks of GRBs. This possibility was studied {by} \citep{viaaxion2000, mirror2004,nucleonbrem2005}, and {in the present study} we compute the profiles {for the} density, temperature, and thickness of the accretion disk by taking into account the cooling rates of all the relevant mechanisms, including advection, neutrino emission, and heavy axion production as the new ingredient. We {consider} the effects {of the non-zero} masses of the axions and pions, following the approach {described by} \cite{nucleonbrem2005}, and {we then} find the ranges of values for the coupling constant $g_{aN}$ for which axions can escape from the disk. If heavy axions decay close to the central engine $d_{a}<10^{9}{\rm cm}${, then} it is possible that a fireball is formed and {this comprises} a mechanism for powering the GRB, as discussed in {the} works mentioned. We shall also explore the possibility that the decays take place further away from the central engine, and leaving the bursts to be powered by a magnetohydrodynamical process (e.g. Blandford-Znajek \cite{bz1977,lee1999}) or by {neutrino--antineutrino} annihilation \citep{pwf1999,zalamea2011}. In this case, the heavy axions produced are still found to leave signatures which are not observed, {because} the energy dependence of the prompt spectrum would be different {compared with that detected, e.g.,} if heavy axions decay preferably to photons. If they decay to electron--positron pairs, {then} the latter can produce a flux of gamma rays {via} IC interactions on the CMB, and this would be observed on Earth, {although} with a very different spectrum {compared with} the typical ones of GRBs.


{The remainder of this} work is organized as follows. In Section 2, we present the calculation of the accretion disk structure and we discuss on the values of the coupling constant $g_{aN}$ for which heavy axions can escape from the disk for different accretions rates. {In} Section 3, we compute the contributions to the gamma-ray flux that would arise from the decaying axions and we compare them with a typical GRB photon flux. Finally, we conclude with a brief discussion {in Section 4}.

\section{Structure of GRB accretion disks with heavy axion production}

In both short and long GRBs, it is expected that matter falls to the newly formed Kerr black hole through a transient, hot and dense accretion disk \cite{pwf1999,kohri2002,dimatteo2002,kohri2005,chen2007}. In {these} disks, energy can be efficiently liberated by advection and via neutrino emission, and the corresponding profiles {for} temperature, density and thickness (or scale height) can be obtained {as} a good approximation using steady state models \cite{dimatteo2002,chen2007,reynoso2006,janiuk2010,nonzero2016}. In this study, we expand the model presented previously \cite{reynoso2006} to include the new process {of interest comprising} the production of heavy axion-like particles. {Similarly to previous studies} \citep{janiuk2010,nonzero2016,janiuk2014}, {we use the notation of Riffert and Herold \cite{riffert1999} for the correcting factors {to} account for general relativistic effects due to the rotating black hole with mass $M_{\rm bh}$ and dimensionless spin parameter $a_*$}:
\be
A&=& 1-\frac{2GM_{\rm bh}}{rc^2 }+\left(\frac{GM_{\rm bh}a_*}{rc^2 }\right)^2 \\
B&=& 1-\frac{3GM_{\rm bh}}{rc^2 }+2a_*\left(\frac{GM_{\rm bh}a_*}{rc^2 }\right)^{3/2}\\
C&=& 1-4a_*\left(\frac{GM_{\rm bh}}{rc^2 }\right)^{3/2}+3\left(\frac{GM_{\rm bh}a_*}{rc^2 }\right)^{2} \\
D&=&\int_{r_{\rm ms}}^{r} \frac{\frac{xc^4}{8G^2}-\frac{3xM_{\rm bh}c^2}{4G}-\frac{3a_*^2M_{\rm bh}^2}{8}}
   {\frac{\sqrt{rx}}{4}\left(\frac{x^2c^4}{G^2}-\frac{3xM_{\rm bh}c^2}{G}+ 2\sqrt{\frac{a_*^2M_{\rm bh}^3c^2x}{G}}\right)} dx,
\ee
{where $r$ is the radius in cylindrical coordinates.}
{These factors were derived from the conservation equation of the {energy--momentum} tensor corresponding to a viscous flow in a spacetime described by the Kerr metric. They are useful {for correcting} the standard expressions for the viscous shear, disk thickness, and heating rate of a Keplerian accretion disk, in order to make them valid for a disk around a rotating black hole. The Keplerian disk case is then recovered if these factors are set to one (see \citep{riffert1999} for details).}

The accretion rate in the disk is supposed to be constant ($\sim 0.1 - 10 \, M_\odot {\rm s}^{-1}$) and mass conservation at a radius $r$ from the black hole implies that  
\beq
\dot{M}= -2\pi v_r\Sigma 
\eeq
where $\Sigma= 2\rho H$ is the mass surface density, $\rho$ is the disk mass density, $v_r$ is the radial velocity and $H$ is the half thickness of the disk. The latter can be written as \citep{riffert1999}
\beq
H\simeq \sqrt{\frac{P r^3}{\rho G M_{\rm bh}}}\sqrt{\frac{B}{C}},\label{eqH}
\eeq
and it is related to the viscous shear as
\beq
f_\phi= \alpha P \frac{A}{\sqrt{BC}}= \frac{\dot{M}}{4\pi H}\sqrt{\frac{GM_{\rm bh}}{r^3}}\frac{D}{A},\label{eqfphi}
\eeq
where $\alpha$ is the {Shakura--Sunyaev} viscosity coefficient \cite{shakurasunyaev1973}. The total pressure is given by  
\beq
P=\rho\frac{kT}{m_n}\left(\frac{1+3X_{\rm nuc}}{4}\right)+ \frac{11}{12}aT^4+ \frac{2\pi hc}{3} \left( \frac{3\rho}{16\pi m_n}  \right)^\frac{4}{3} + \frac{u_\nu}{3},\label{eqP}
\eeq
where the fraction of free nuclei is approximated by \citep{kohri2005}
$$ X_{\rm nuc}\approx 295.5 \rho_{10}^{ -\frac{3}{4} }T_{11}^{ \frac{9}{8} } \exp^{-0.82/T_{11}}, $$
with $\rho_{10}=\frac{\rho}{ 10^{10} {\rm g \ cm^{-3}} }$ and $T_{11}=\frac{T}{10^{11}{K}}$. Electron neutrinos and antineutrinos are the ones which are {produced more efficiently and they} can become trapped, {thereby} contributing to the pressure. To describe their energy density we follow Ref. \citep{dimatteo2002}  and adopt the prescription:
$$	
u_{\nu}= \frac{7}{8}\sigma T^4  \sum_{l=\{e,\mu,\tau \}} \left(\frac{\frac{\tau_{\nu_l}}{2}+ \frac{1}{\sqrt{3}}}  
{\frac{\tau_{\nu_l}}{2}+ \frac{1}{\sqrt{3}} +\frac{1}{3\tau_{a,\nu_l}} } \right),
$$
where the neutrino optical depth is the sum of the scattering plus the absorptive contributions, $\tau_{\nu_l}=\tau_{s,\nu_l}+\tau_{a,\nu_l}$, which are given {in the following}.

 {Accretion proceeds} as the energy generated by friction is either advected {toward} the black hole or emitted by the disk. The heating rate due to viscosity can be written as
\beq
{Q}^+_{\rm vis}= \frac{3\dot{M}GM_{\rm bh}}{8\pi H r^3}\frac{D}{B}.\label{eqnQvis}
\eeq 
The steady-state solutions {are obtained} requiring that the heating rate is equal to the total cooling rate at each radius, ${Q}^-=  {Q}^+_{\rm vis}$, including all the relevant cooling processes:
\beq
{Q}^{-}= {Q}^{-}_{\rm phot}+ {Q}^{-}_{\rm adv}+ {Q}^{-}_{\nu}+ {Q}^{-}_{a}.  \label{eqnQcool}
\eeq
The rate of photo-disintegration {for} heavy nuclei is  given by \cite{pwf1999}
\beq
Q_{\rm phot}=10^{29} { \rho_{10} }v_r \frac{dX_{\rm nuc}}{dr}H,
\eeq
and the cooling by advection can be approximated as \cite{narayanyi1994}
\beq
{Q}^{-}_{\rm adv}\simeq \frac{v_r}{r}\left[\frac{38}{9}aT^4+\frac{3\rho kT}{8m_N}\left( 1+X_{\rm nuc}\right)\right].
\eeq
Neutrino cooling occurs mainly through the {electron--positron} pair capture process, $p+e^-\rightarrow n+\nu_e$ and $n+e^+\rightarrow p+\bar{\nu}_e$, at a rate
\beq
{Q}^{-}_{Ne\rightarrow \nu_e\bar{\nu}_e}= 9.2\times 10^{33}\rho_{10}T_{11}^6 X_{\rm nuc} {\rm erg \ cm^{-3}s^{-1}},
\eeq
and also through {electron--positron} pair annihilation at rates
\be
{Q}^{-}_{e^+e^-\rightarrow \nu_e\bar{\nu}_e} &=& 3.4\times 10^{33}T_{11}^9	{\rm erg \ cm^{-3}s^{-1}},\\ {Q}^{-}_{e^+e^-\rightarrow \nu_\mu\bar{\nu}_\mu} &=& {Q}^{-}_{e^+e^-\rightarrow \nu_\tau\bar{\nu}_\tau}= 0.7\times 10^{33}T_{11}^9	{\rm erg \ cm^{-3}s^{-1}}.
\ee

{Considering} the corresponding inverse processes, the absorption optical depth can be approximated as
\beq
\tau_{a,\nu_l}\approx \frac{{Q}^{-}_{\nu} H}{4\frac{7}{8}\sigma T^4},  
\eeq
and the scattering {process} as  $\tau_{s}= 2.7\times 10^{-7}T_{11}^2 H$. Then, {by using} a simplified treatment for neutrino emission and transport based on a two-stream approximation (e.g., {see} \citep{dimatteo2002,hubeny1990}), we obtain the escaping energy rate in neutrinos as {follows:}
\beq
{Q}^{-}_{\nu}= \sum_{l=\left\{e,\mu,\tau \right\}} \frac{\frac{7}{8}\sigma T^4} {\frac{3}{4}\left[\tau_{\nu_l}/2 +1/\sqrt{3}  
+1/(3\tau_{a,\nu_l})\right]}.
\eeq 


Finally, we include the energy loss rate due to the emission of heavy axions {by {nucleon--nucleon} bremsstrahlung} according to \citep{nucleonbrem2005}, without neglecting the finite mass of the axions and pions, 
\beq  
{Q}^{-}_{a}=  \int_{{m_ac^2}}^\infty E_a I_a(E_a) dE_a,
\eeq
where $I_a(E_a)=\frac{d\mathcal{N}_a}{dE_a}$ is the intensity of the emitted axions,
\begin{multline} 
I_a(E_a)=\frac{35}{28} \frac{\tilde{Q}_a}{(kT)^2} \left[1-\frac{m_a^2c^4}{E_a^2}\right]^{\frac{3}{2}}
  \\ \int_0^\infty dv \sqrt{ v^2 + v\frac{E_a}{kT} } e^{\left(-v-\frac{E_a}{kT}\right)} \int_{-1}^1 \eta dz,  
\end{multline}
{with}
\beq 
  \tilde{Q}_a= 3.4\times 10^{42} T_{\rm MeV}^\frac{7}{2} \rho_{12}^2 g_{aN}^2 \ {\rm erg \ cm^{-3} \ s^{-1}}
\eeq 
and $\eta= \left(\eta_{\vec{k}}+\eta_{\vec{l}}+\eta_{\vec{k}\vec{l}}-3\eta_{\vec{k}\cdot\vec{l}}\right)$, where
\be
\eta_{\vec{k}}&=& \left.\left(\frac{u+v-2z\sqrt{uv}}{u+v-2z\sqrt{uv}+\frac{m_\pi^2c^2}{m_N kT}}\right)^2\right|_{u=v+\frac{E_a}{kT}} \\
\eta_{\vec{l}}&=&\left. \left(\frac{u+v+2z\sqrt{uv}}{u+v+2z\sqrt{uv}+\frac{m_\pi^2c^2}{m_N kT}}\right)^2\right|_{u=v+\frac{E_a}{kT}} \\
\eta_{\vec{k}\vec{l}}&=&\left. \frac{(u+v)^2-4uvz}{\left(u+v+\frac{m_\pi^2c^2}{m_N kT}\right)^2-4uvz}\right|_{u=v+\frac{E_a}{kT}} \\
\eta_{\vec{k}\cdot\vec{l}}&=&\left. \frac{(u-v)^2}{\left(u+v+\frac{m_\pi^2c^2}{m_N kT}\right)^2-4uvz}\right|_{u=v+\frac{E_a}{kT}}.
\ee

It is also useful to compute the mean free path for the inverse process, {i.e.,} axion capture by nucleons, which can be obtained as \citep{nucleonbrem2005}
\begin{multline}
\lambda_a^{-1}(E_a)= 8.75\times 10^{-5}{\rm cm^{-1}}\frac{\tilde{Q}_a }{(kT)^3E_a}\left[1-\frac{m_a^2c^4}{E_a^2}\right]^{\frac{1}{2}}			\\
\int_0^\infty dv \sqrt{(v+x)v}e^{-v}\int_{-1}^1 \frac{\eta}{2}dz.	
\end{multline}

\begin{figure}[tbp]
\centering 
\includegraphics[width=.5\textwidth,trim=0 0 400 0,clip]{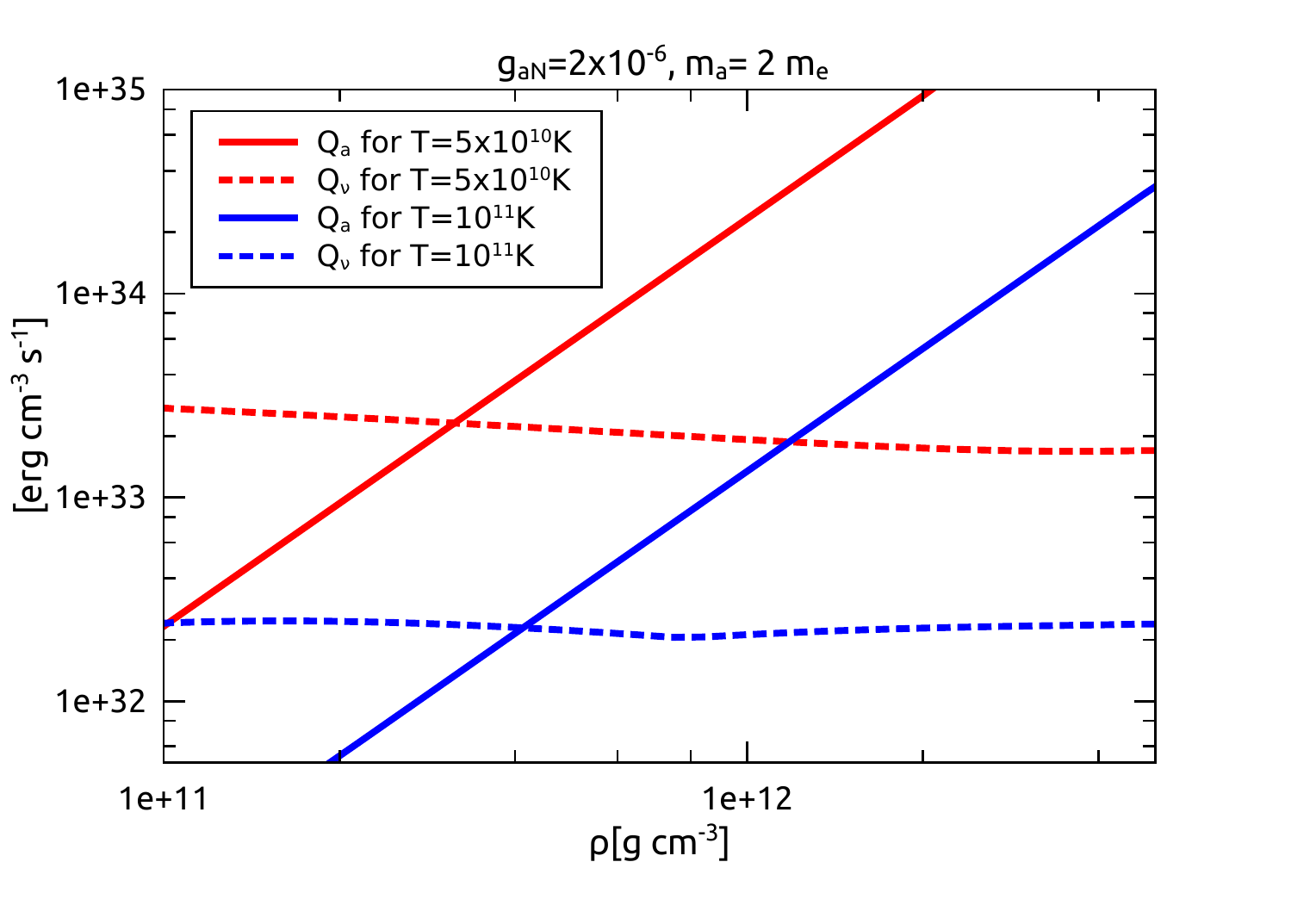}
\caption{\label{fig.qaxqnu} Cooling rates for axion bremsstrahlung and neutrino emission.}
\end{figure}

{Figure \ref{fig.qaxqnu} shows} the obtained cooling rate $Q_a$ as a function of the density for $T=5\times 10^{10}K$ and $T=10^{11}$K for $g_{aN}=2\times 10^{-6}$ and $m_a= 2 m_e c^2$ as compared to the corresponding neutrino cooling rates. It can be seen from this plot that axion emission becomes more important than neutrino cooling for densities $\rho \gtrsim 3 \times 10^{11} {\rm g \ cm^{-3}}$ {at the temperatures} shown, which are typical values for the central part of {the} accretion disks in GRBs. 

In order to calculate the density, thickness, and temperature of the disk as {functions} of the radius $r$, we proceed as {described by} \cite{reynoso2006} and first solve numerically Eqs. (\ref{eqH}) and (\ref{eqfphi}) to obtain $H$ and $P$ as {functions} of the density $\rho$. Then, {by using} of the equation of state (Eq. \ref{eqP}), we can {also} obtain $T$ as a function of $\rho$, and {we employ this relation to} evaluate the total cooling rate (Eq. \ref{eqnQcool}) and equate it to the heating rate (Eq. \ref{eqnQvis}
), {thereby obtaining} the correct pair $(\rho,T)$ that satisfies the energy balance at each radius $r$ .   

\begin{table}[tbp]
\centering
\begin{tabular}{|lc|c|}
\hline
Parameter & Description & Values\\
\hline
$M_{\rm bh}$ & black hole mass &  $3 M_\odot$ \\
$a_*$ & black hole spin & $0.9$ \\
$\alpha$ & viscosity parameter  & $0.1$\\
$\dot{M}$ & accretion rate & $\left\{0.1, 1, 3 \right\}M_\odot {\rm s^{-1}}$\\
$g_{aN}$ & {axion--nucleon} coupling & $5\times 10^{-7}-10^{-5}$\\
$m_a$ & heavy axion mass  &  $\left\lbrace 0.1,1.022, 10\right\rbrace {\rm MeV} $\\
\hline
\end{tabular}
\caption{\label{table.params} Parameters used in the accretion disk model with heavy axion production.}
\end{table}

\begin{figure*}[tbp]
\includegraphics[width=.9\textwidth,trim=0 450 650 0,clip]{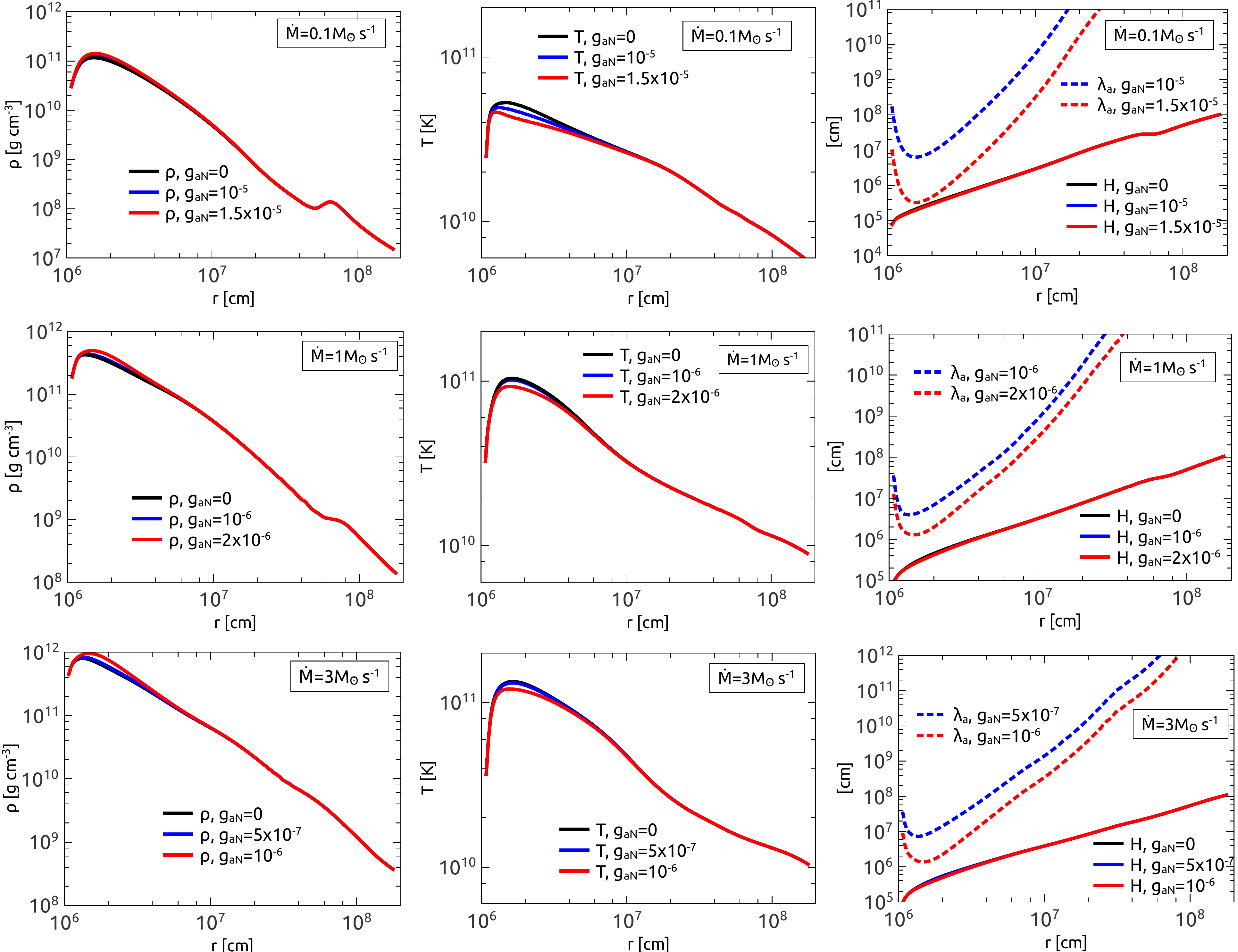}
\caption{\label{fig.structure} Density (left panels), temperature (middle panels), {and} thickness and axion mean free path (right panels). In the top panels, $\dot{M}=0.1\, M_\odot {\rm s^{-1}}$ and $g_{aN}=\left\lbrace 10^{-5},1.5\times 10^{-5} \right\rbrace$. In the middle panels, $\dot{M}=1\, M_\odot {\rm s^{-1}}$ and $g_{aN}=\left\lbrace 10^{-6},2\times 10^{-6} \right\rbrace$. {In} the bottom panels, $\dot{M}=3\, M_\odot {\rm s^{-1}}$ and $g_{aN}=\left\lbrace 5\times 10^{-7},10^{-6} \right\rbrace$.}
\end{figure*}

{Figure \ref{fig.structure} shows the results obtained} for the profiles $\rho(r)$, $T(r)$, and $H(r)$ in the left, middle, and right panels, respectively. The values of the parameters {used} are summarized in Table \ref{table.params}. {The accretion rate is $\dot{M}=0.1 \, M_\odot{\rm s^{-1}}$ in the top panels, $\dot{M}=1 \, M_\odot{\rm s^{-1}}$ in the center panels, and $\dot{M}=3 \, M_\odot{\rm s^{-1}}$ in the bottom panels}. {In this plot, we assume} that the mass of the heavy axions is $m_a= 2m_e$. {These results show that} the density, temperature, and thickness do not change significantly with respect to the case {with} no axion production, at least for the {coupling strength ranges considered}. {In the left panels, we also show} the mean free path for the axions ($\lambda_a$) evaluated at their mean energy (which turns out to be $E_a\simeq 2 kT$ \cite{nucleonbrem2005}), {compared with} the thickness $H(r)$. Hence, we can conclude that for couplings as high as $g_{aN}\simeq 10^{-5}$, heavy axions with mass $m_a= 2 m_e$ will escape freely from the disk if $\dot{M}\lesssim 0.1 M_\odot {\rm s^{-1}}$, {whereas for} disks with higher accretion rates $\dot{M}= 1-3 M_\odot {\rm s^{-1}}$, {axions will} escape without interacting for couplings $g_{aN}\lesssim 10^{-6}$. For higher values of $g_{aN}$, axions will become trapped in the disk but we do not address such cases in the present {study}.

\begin{figure}[tbp]
\centering 
\includegraphics[width=.75\textwidth,trim=0 20 180 0,clip]{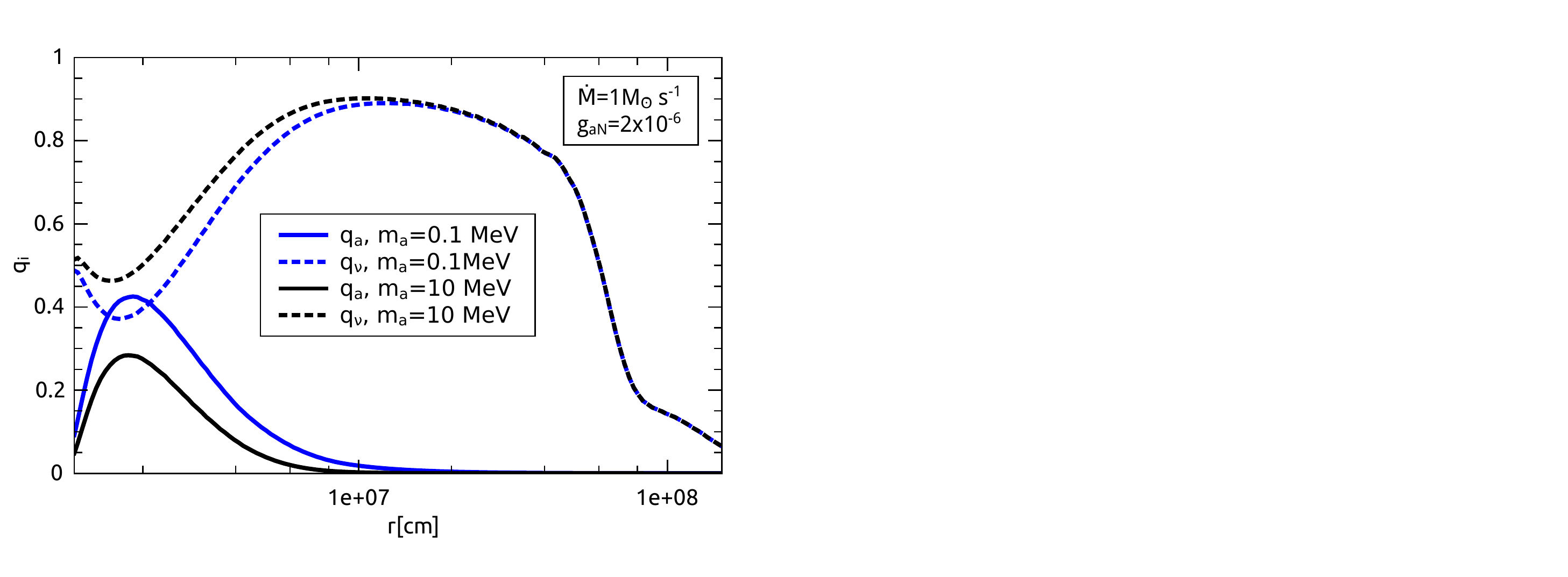}
\caption{\label{fig.fnufax} $q_{\nu}$ and $q_a$ as a function of the disk radius. }
\end{figure}

For different values of the heavy axion mass, their emission from the disk is also different, and in particular less intense for higher masses. This can be seen in Figure \ref{fig.fnufax}, where we plot the relative cooling parameter for axions $q_{a}= Q^-_{a}/Q^-_{\rm tot}$ and neutrinos  $q_\nu= Q^-_{\nu}/Q^-_{\rm tot}$ {as functions} of the disk radius in the case of an accretion rate $\dot{M}=1 M_\odot {\rm s}^{-1}$ for a coupling $g_{aN}=2\times 10^{-6}$, and for heavy axion masses $m_a= \left\lbrace 0.1 {\rm MeV}, 10 {\rm MeV}\right\rbrace$. {It can be seen from this plot that at the innermost regions of the disk, neutrino cooling becomes less efficient than at its maximum values {reached further away from the center}. This is due {to increased} neutrino trapping,  which implies that the advection process becomes more significant, {thereby} leading to a decrease in the density and the temperature in agreement with the results {by} \citep{chen2007,nonzero2016}. }


\section{Heavy axion emission from GRB accretion disks: Implications}

In this section, we study what is to be expected as observational consequences in the case that heavy axions are produced in GRB disks {as described above}. Again, as mentioned {by} \citep{viaaxion2000,mirror2004,nucleonbrem2005}, if the axions generated in the disk are to decay close enough ($d_a<10^{9}{\rm cm}$), the $e^+e^-$ fireball can be formed more efficiently than by neutrino-antineutrino annihilation. {The decay rates of these} heavy axions to photons and to {electron--positron} pairs are
\be
\Gamma_{a\rightarrow \gamma\gamma}&=& 1.5\times 10^{21} \frac{g_{a\gamma}^2m^2_a}{64\pi }\left(\frac{m_a c^2}{\rm MeV}\right)  {\rm s^{-1}}\\
\Gamma_{a\rightarrow e^+e^-}&=& 1.5\times 10^{21} \frac{g_{ae}^2}{8\pi}\left(1-\frac{4m_e^2}{m_a^2}\right)^{\frac{1}{2}}\left(\frac{m_a c^2}{\rm MeV}\right)  {\rm s^{-1}},
\ee   
{so we find that}, for instance, for axions of mass $m_a\sim 1{\rm MeV}$, the coupling constants {must} be $g_{a\gamma}\gtrsim 10^{-5.1}{\rm GeV^{-1}}$ {and/or} $g_{ae}\gtrsim 10^{-8.8}$ for the decays to occur at distances {less} than $10^{9}{\rm cm}$ from the central black hole. {{Under these} conditions, the standard phenomenology of the fireball model can then be {employed} to describe the burst, i.e., the created pairs or photons would form {an} optically thick plasma, {which will} expand {due to} radiation pressure and generate gamma-rays {via} internal shocks at larger distances where the flow becomes optically thin \cite{meszaros2006,piran1999}.} 

{We} also consider the possibility that the decays can take place far from the central engine, {such as} in the interstellar medium or even outside the host galaxy. In this case, the GRBs {will not} be powered by axion decay {but by other mechanism instead,}  such as neutrino--antineutrino annihilation or a magnetohydrodynamic mechanism (e.g. Blandford--Znajek). The high luminosity of the emitted axions, implies a high luminosity of the decay products, which would be directly observable in the case of photons, or {via} the IC radiation generated by the produced $e^+e^-$ scattering on the CMB.

As mentioned {in previous studies} \citep{viaaxion2000,mirror2004}, axions would escape freely from short GRB accretion disks generated in compact merger events. {Furthermore,} it can be seen that they would also be {capable of escaping} from the collapsing star in long GRBs. {To estimate} the corresponding optical depth, we consider that the central temperature of a GRB progenitor is $T_{0} \lesssim 10^{10}{\rm K}$ and the density is $\rho_{0}\lesssim 10^{10}{\rm g \ cm^{-3}}$ (e.g. \cite{woosley2002}). If we assume that the density drops on the envelope as $\rho_{\rm e} =\rho_0\left(\frac{r}{r_0}\right)^{-1.5}$  (e.g. \cite{meszarosmurase2016}), {then by} taking $r_0=10^4{\rm cm}$, we find that the optical depth for axions of energies $\sim 1 \, {\rm MeV}$ is much less than one in {all of the} cases studied:  
\beq
\tau_a < \int_{r_0} \frac{dr_{\rm e}}{\lambda_a(T_0,\rho_{\rm e}(r_{\rm e}))}< 10^{-6}, 
\eeq
{and thus} in the present context, we can consider that {also} axions escape freely from the stellar envelope in long GRBs.

The luminosity of the emitted axions can be calculated as
\beq
L_a=  2\pi\, \int_{r_{\rm ms}}^{R_{\rm out}} dr \, r\, 2\,H(r) \rho(r)Q_a(r), 
\eeq
and an analogous expression is used for the neutrino luminosity $L_\nu$. In the {expression above}, $R_{\rm out}=200\, r_{\rm g}$ is the outer radius of the disk and the inner radius is taken as {that} corresponding to the last stable circular orbit,
$$r_{\rm ms}=r_{\rm g}\left[3+z_2- \sqrt{(3-z_1)(3+z_1+2z_2)}\right],$$
where $r_g= 2GM_{\rm bh}/c^2$, $z_1=1+(1-a_*^2)^{\frac{1}{3}}\left[(1+a_*)^\frac{1}{3}+ (1- a_*)^\frac{1}{3} \right]$, and $z_2=\left(3a^2+z_1^2 \right)^\frac{1}{2}$.

We can also estimate the power that would be generated by the Blandford--Znajek process using the expression given {by} \cite{xue2013}:
\beq
L_{\rm BZ} \approx c r_g^2 \frac{B_{\rm in}^2}{8\pi}, 
\eeq
where $B_{\rm in}$ is the poloidal magnetic field near the horizon. The latter can be related to the pressure {via} $B^2_{\rm in}/(8\pi)\approx P_{\rm in}$, {and the pressure is} approximated by (see \cite{liu2015}) $P_{\rm in}\approx 10^{ \left[ 30 + 1.22\, a_\star+ \log(\dot{M}/(M_\odot {\rm s^{-1}})) \right] }{\rm erg \, cm^{-3}} $. {For comparison, Table \ref{table.lum} shows} the obtained values for $L_a$, $L_\nu$, and $L_{\rm BZ}$ {using} different values of the accretion rate, mass of the heavy axion, and their coupling constant to nucleons. It is important to {note} that the efficiency of neutrino--antineutrino annihilations is always less than $10\%$ in the cases studied of black hole spin and accretion rates, as pointed out {by} \cite{liu2015}. In general this process is less efficient than the {Blandford--Znajek process}, which varies here only with the accretion rate {because} we keep $a_\star=0.9$ in all cases. The luminosity in heavy axions clearly increases with $g_{aN}$ and with the accretion rate, since the latter implies a higher density and temperature.

\begin{table*}[tbp]
\centering
\begin{tabular}{|l|c|c|c|}
\hline
$(\dot{M}[M_\odot{\rm s^{-1}}], m_a[{\rm MeV}], g_{aN})$ & $L_a[{\rm erg \ s^{-1}}] $ & $L_\nu[{\rm erg \ s^{-1}}]$ & $L_{\rm BZ}[{\rm erg \ s^{-1}}]$ \\
\hline
$(0.1, \ 1, \  10^{-5})$ & $4.1\times 10^{51}$ &  $1.9\times 10^{52}$ & $7.4\times 10^{51}$ \\
$(0.1, \ 1, \ 1.5\times 10^{-5})$ & $7.6\times 10^{51}$ &  $1.5\times 10^{52}$ & $7.4\times 10^{51}$ \\
$(1, \ 0.1, \ 10^{-6})$ & $1.3\times 10^{52}$ &  $2.1\times 10^{53}$ & $7.4\times 10^{52}$ \\
$(1, \ 10, \ 10^{-6})$ & $7.4\times 10^{51}$ &  $2.2\times 10^{53}$ & $7.4\times 10^{52}$ \\
$(1, \ 0.1, \ 2\times 10^{-6})$ & $4.9\times 10^{52}$ &  $1.9\times 10^{53}$ & $7.4\times 10^{52}$ \\
$(1, \ 10, \ 2\times 10^{-6})$ & $2.8\times 10^{52}$ &  $2.0\times 10^{53}$ & $7.4\times 10^{52}$ \\
$(3, \ 0.1, \ 5\times 10^{-7})$ & $3.8\times 10^{52}$ &  $5.5\times 10^{53}$ & $2.2\times 10^{53}$ \\
$(3, \ 10, \ 5\times 10^{-7})$ & $2.6\times 10^{52}$ &  $5.6\times 10^{53}$ & $2.2\times 10^{53}$ \\
$(3, \ 0.1, \ 10^{-6})$ & $1.6\times 10^{53}$ &  $4.7\times 10^{53}$ & $2.2\times 10^{53}$ \\
$(3, \ 10, \  10^{-6})$ &  $10^{53}$ &  $5.1\times 10^{53}$ & $2.2\times 10^{53}$ \\
\hline
\end{tabular}
\caption{\label{table.lum} Luminosity in heavy axions, neutrinos, and due to the Blandford--Znajek process.}
\end{table*}

Without {considering the specific details of} how the prompt emission is generated in GRBs, we can still rely on observations {of} the energy dependence of the detected gamma-ray flux and the observed luminosity. A standard fit to data on many bursts is the so-called Band flux \cite{band1993}  
\beq 
\phi^{\rm Band}_\gamma(E_\gamma)= A
\left\{
  \begin{array}{c}
   \left(\frac{E_\gamma}{100{\rm keV}}\right) ^{\alpha_0} e^{-\frac{E_\gamma}{E_0}}  \ {\rm for} \ E_\gamma\leq E_0 (\alpha_0-\beta_0) \\
  \left( \frac{E_0(\alpha_0-\beta_0)}{100{\rm keV}} \right)^{\alpha_0-\beta_0} \left(\frac{E_\gamma}{100{\rm keV}}\right) ^{\beta_0} e ^{(\alpha_0-\beta_0)\frac{E_\gamma}{100{\rm keV}}}  \ {\rm otherwise}, 
  \end{array} 
  \right.
\eeq
with $E_0=100 {\rm keV}$, and for the low- and high-energy indexes{,} we take $\alpha_{0}=-1$, and $\beta_0=2.1$ based on the analysis of {the Fermi collaboration using} four years of data taking on many bursts with the Gamma-ray Burst Monitor \cite{fermigbm2014}. The constant $A$ can be fixed by normalization on the total flux in a given energy band, and we consider the band $(10 {\rm keV}-40 {\rm MeV})$ and a flux $\Phi_\gamma\approx 2.5 {\ \rm ph \ cm^{-2} s^{-1}}$ which according to Ref. \cite{fermigbm2014}, is  close to the mean flux according to their analysis on the samples. 

\begin{figure*}[tbp]
\centering 
	\includegraphics[width=.9\textwidth,trim=10 35 40 50,clip]{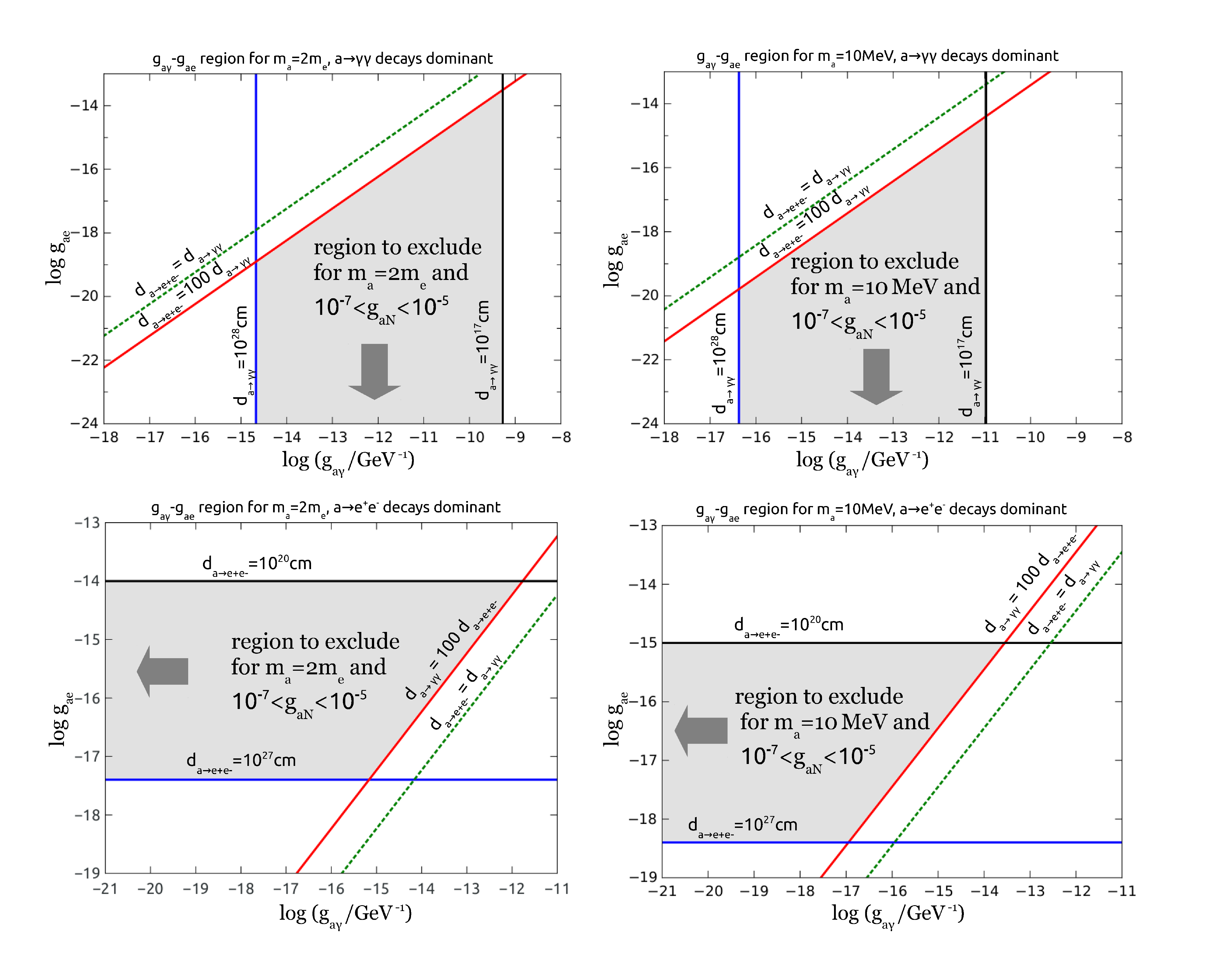}
\caption{\label{fig.shaded} Regions in the $g_{a\gamma}-g_{ae}$ plane that correspond to dominant ${a\rightarrow \gamma\gamma}$ decays in the top panels, and to dominant $a\rightarrow e^+e^-$ decays in the lower panels, {and} for $m_a=2m_e$ and $m_a=10\, {\rm MeV}$ on the right and left plots, respectively. The shaded regions can be excluded if $10^{-7}\lesssim g_{aN}\lesssim 10^{-5}$, for which {axions} can be efficiently emitted from GRB accretion disks. The arrows {denote} that the regions can be extended arbitrarily in the  {directions indicated}.}
\end{figure*}

We then take a Band flux at the {level mentioned above} and compare it to the fluxes that would arise from the decay of heavy axions if they decay far from the central engine and before arriving on Earth. {In particular}, we consider the possibility that they decay primarily to photons at distances $d_{a}= \frac{E_a}{m_a c^2}\frac{v_a}{\Gamma_{a\rightarrow \gamma\gamma}} $, such that $10^{17}{\rm cm}<d_{a}<10^{28}{\rm cm}$ for GRBs at redshift $z=1$, i.e. at a luminosity distance $d_L= 3.6\times 10^{28}{\rm cm}$. We {consider} this particular distance for illustration because the distribution of GRB with redshift {exhibits} its higher values for $z\sim 1$ \cite{fermigbm2014}. {Given} the above expressions for the decay rates,
{we can see} that $\Gamma_{a\rightarrow\gamma\gamma}= \Gamma_{a\rightarrow e^+e^-}$ {will hold, e.g.,} if $g_{ae}= (5.8\times 10^{-4}{\rm GeV})g_{a\gamma} $ for $m_a=2m_e$, and if $g_{ae}= (3.5\times 10^{-3}{\rm GeV})g_{a\gamma} $ for $m_a=10\, {\rm MeV}$. Hence,  
the ranges of the couplings $g_{a\gamma}$ and $g_{ae}$ that correspond to the channel $a\rightarrow \gamma\gamma$ as dominant are marked {by the shaded regions in }the upper plots in Fig. \ref{fig.shaded}, for $m_a=2m_e$ and $m_a=10\,{\rm MeV}$ in the left and right panels, respectively.

For simplicity, the flux of gamma rays produced in this case can be estimated assuming an isotropic emission as
\beq
  \phi_{a\rightarrow \gamma\gamma}(E_\gamma)= \frac{(1+z)^2}{4\pi d_L^2} \int_{r_{\rm ms}}^{r_{\rm out}}dr 4\,H(r) \, r \int_{E_a^{\rm min}}^\infty \frac{dE_a 2}{\sqrt{{E_a}^2- m_a^2c^4}}I_a(E_a),
\eeq 
where $E_a^{\rm min}= \frac{E'_\gamma}{2}+ \frac{m_a^2c^4}{2\,E_\gamma'}$ and $E_\gamma'=E_\gamma (1+z)$.

On the other hand, if the dominant decay channel is $a\rightarrow e^+e^-$, {then} we consider decay lengths such that $10^{20}{\rm cm}<d_{a}<3\times10^{26}{\rm cm}$, i.e., {up to a maximum of} $400 {\rm Mpc}$ from the burst site, in order to consider that the interactions with the CMB are initiated at $z\simeq 1$ but not closer than $\sim 30\,{\rm kpc}$ to avoid $e^{\pm}$ {remaining} trapped in the magnetic field of the host galaxy. The shaded regions {in} the bottom plots in Fig. \ref{fig.shaded} indicate the ranges of the couplings $g_{ae}$ and $g_{a\gamma}$ {at which these} decays are dominant, for $m_a=2m_e$ and $m_a=10\,{\rm MeV}$ in the left and right panels, respectively.

In order to estimate the flux of scattered photons that would arrive on Earth, we follow the analytical treatment of electromagnetic cascades {described by} \citep{venyakala2016}, such that the spectrum of cascade photons initiated by an electron of energy $E_e$ can be {generally expressed} as 
\be
\frac{dn_{\rm cas}(E_e,E_\gamma)}{dE_\gamma}
\left\lbrace
\begin{array}{cc}
\frac{K}{\mathcal{E}_{X}} \left(\frac{E_\gamma}{\mathcal{E}_X}\right)^{-1.5} \ \ {\rm for} \ \ E_\gamma< \mathcal{E}_X \\
\frac{K}{\mathcal{E}_{X}} \left(\frac{E_\gamma}{\mathcal{E}_X}\right)^{-2} \ \ {\rm for} \ \ \mathcal{E}_X<E_\gamma<    \mathcal{E}_\gamma \\
0  \ \ {\rm for} E_{\gamma}>\mathcal{E}_\gamma.
\end{array} 
\right.
\ee
Here, $\mathcal{E}_\gamma= m_e c^2/[\epsilon_{\rm ebl}(1+z)] $, $\mathcal{E}_X=\frac{1}{3}\left(\mathcal{E}_\gamma/m_ec^2\right)^2\epsilon_{\rm cmb}(1+z)$, and the characteristic energies of the CMB and the EBL are $\epsilon_{\rm cmb}=6.3 \times 10^{-4}{\rm eV}$ and $\epsilon_{\rm ebl}=0.68 \, {\rm eV}$, respectively . We adopt this dichromatic approximation for the background photons, although the $e^\pm$ produced will interact mainly with the CMB radiation and the scattered photons are cascade sterile, {because }they are below the threshold for $e^+e^-$ production. The parameter $K$ can be computed by considering energy conservation between the initial $e^\pm$ energy and the total energy of the scattered photons. We note that in the present context and for redshifts $z\lesssim 2$, the break energy is $\mathcal{E}_X\gtrsim 15 \,{\rm MeV}$, so that most $e^\pm$ have energies $E_e<\mathcal{E}_X$, and hence the scattered photons have a spectrum $\propto E_{\gamma}^{-\frac{3}{2}}$. In these cases, it is found that $K= \frac{1}{2}\sqrt{E_e/\mathcal{E}_X}$, {whereas} if $\mathcal{E}_e<E_e<\mathcal{E}_\gamma$, then $K=E_e/\left[\mathcal{E}_X(2+\ln(E_e/\mathcal{E}_X))\right]$. {According to \cite{venyakala2016}, if we assume} that the interactions occur in less than a timescale $\sim H_{\rm hubble}^{-1}(z)$, {then} we can write the flux of IC scattered photons to be observed on Earth as
\beq 
\phi_{e^\pm}^{\rm IC}(E_\gamma)= \frac{(1+z)^2}{4\pi d_L^2}\int_{E_\gamma'}^\infty dE_e I_e(E_e) \frac{dn_{\rm cas}(E_e,E_\gamma')}{dE'_\gamma}.
\eeq  
Here, the intensity of decaying electrons and positrons is given by 
\beq
I_e(E_e)= \int_{E_a^{\rm min}}^\infty dE_a \frac{I_a(E_a)} { \gamma_a\beta_a\sqrt{\frac{m_a^2c^4}{4}-m_e^2c^4} },
\eeq
where $\gamma_a=\frac{E_a}{m_ac^2}$ is the Lorentz factor of the axion and $\beta_a$ {is} its velocity in units of $c$. The minimum energy for the decay to $e^\pm$ of energy $E_e$ is given by 
$$E_{a}^{\rm min}=\frac{m^2_a}{2m_e^2}\left[E_e-\sqrt{E_e^2-4\frac{m_e^2}{m_a^2}\left(E_e^2+\frac{m_a^2c^4}{4}- m_e^2c^4\right)}\right],$$  
which follows {when we consider} that the maximum and minimum $e^\pm$ energies in the laboratory frame add up to $E_a$, similarly to the discussion {by} \cite{stecker1971}.

\begin{figure*}[tbp]
\centering 
	\includegraphics[width=.9\textwidth,trim=5 1260 60 0,clip]{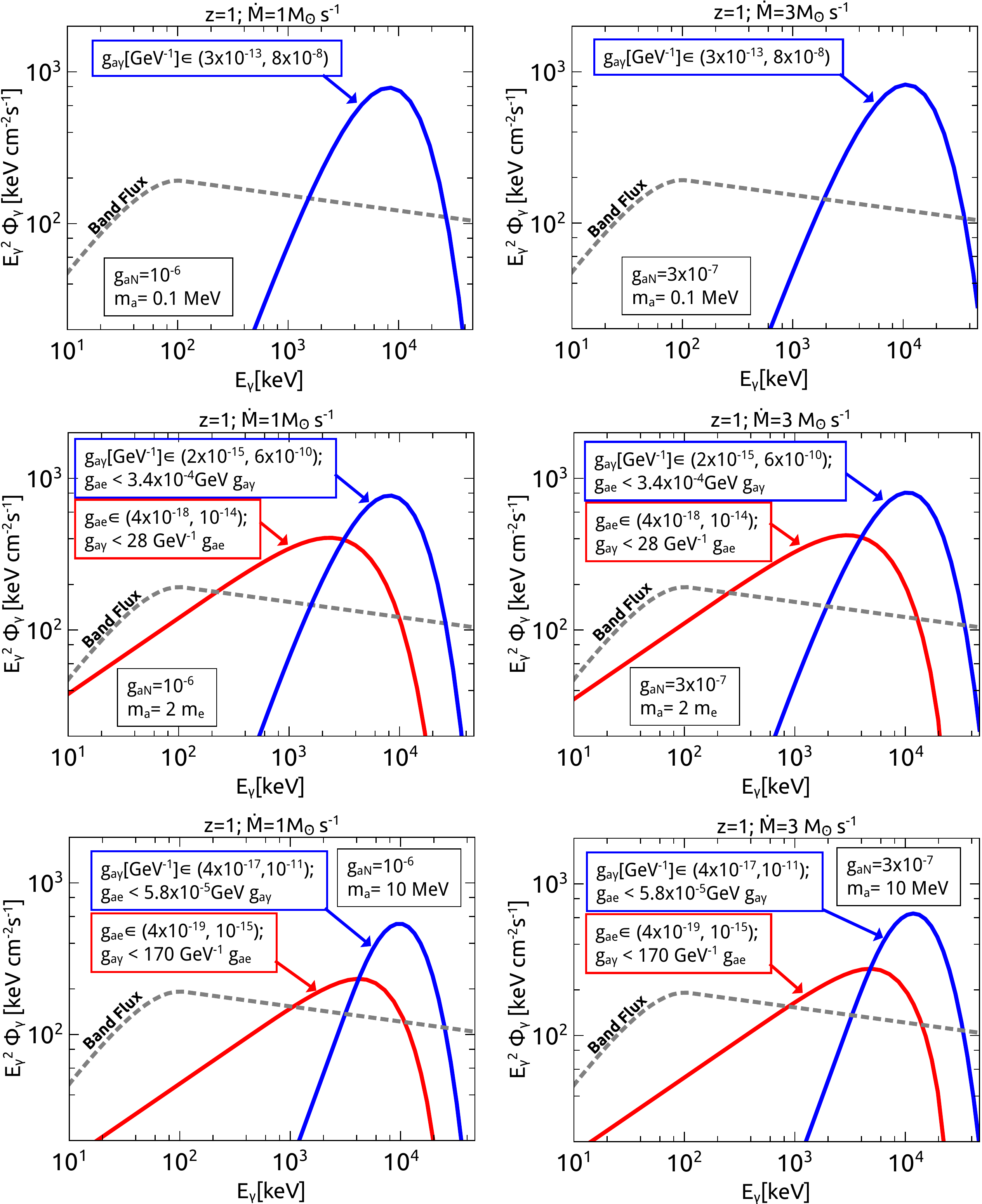}
\caption{\label{fig.fluxes} Contributions to the flux of gamma rays {due to} the decay of heavy axions produced in the accretion disk of a GRB at $z=1$, and for accretion rates $\dot{M}=1M_\odot{\rm s^{-1}}$ in the left panels and $\dot{M}=3M_\odot {\rm s^{-1}}$ in the right panels. The mass of the heavy axions $m_a$ is $0.1$ MeV, $2m_e$, and $10$ MeV in the top, middle and bottom panels, respectively. Blue lines correspond to $\phi_{a\rightarrow \gamma\gamma}(E_\gamma)$ and red lines to $\phi_{e^{\pm}}^{\rm IC}(E_\gamma)$, {and} a typical GRB flux is shown {by} dashed gray curves.}
\end{figure*}

{Figure \ref{fig.fluxes} shows the flux contributions obtained} in gamma rays from the decay of heavy axions in the cases of direct decay to photons and also if the decay to electron--positrons is dominant. In the former case, the flux to be observed would be very different from the typical Band flux, which appears in dashed grey lines, {whereas} the contribution from direct {decays to photons is} in blue.  
We note that even if we considered a Band flux {twice as high as} the level shown in the plots, {the contributions of decaying} axions would {still} introduce visible signatures {into} the spectra. And in this case, it appears reasonable to expect that most of the GRBs presents such a high flux or less ($\Phi_\gamma \lesssim 5 {\ \rm ph \ cm^{-2} s^{-1}}$), according to the analysis of Fermi-GBM ({see Fig. 11 in} \citep{fermigbm2014}). {{Therefore}, the features corresponding to the red and blue curves {in} Fig. \ref{fig.fluxes} would have been observed in most of the GRBs, and {thus} for $10^{-7} \lesssim g_{aN}\lesssim 10^{-5}$, we can exclude the ranges of $g_{a\gamma}$ and $g_{ae}$  marked {by} the shaded regions {in} Fig. \ref{fig.shaded}.}

Now, in the cases of axions {that decay} dominantly to $e^+e^-$, the flux contributions due to IC on the CMB are found to be significant and more spread {widely} to lower energies {(red curves {in} Fig. \ref{fig.fluxes})}. {Actually}, this component of the flux is not expected to arrive from the same direction {as} the original burst emission, which is supposed to be beamed. This is because the electrons and positrons {are deviated} in the intergalactic magnetic field (e.g.\citep{venyakala2016}). {Hence, for coupling values  $10^{-7} \lesssim g_{aN}\lesssim 10^{-5}$ and $(g_{ae},g_{a\gamma})$ within the shaded regions {in} the lower panels of Fig. \ref{fig.shaded}, the mentioned flux component should have been clearly observed, i.e., not superimposed directionally and temporally to a Band-like flux. The lack of observations of any flux contribution {such as the} described allows us to exclude the {aforementioned} combination of values {for} $g_{aN}$, $g_{a\gamma}$, and $g_{ae}$. }

\begin{figure*}[tbp] 
\centering 
		\includegraphics[width=.9\textwidth,trim=0 5 0 0,clip]{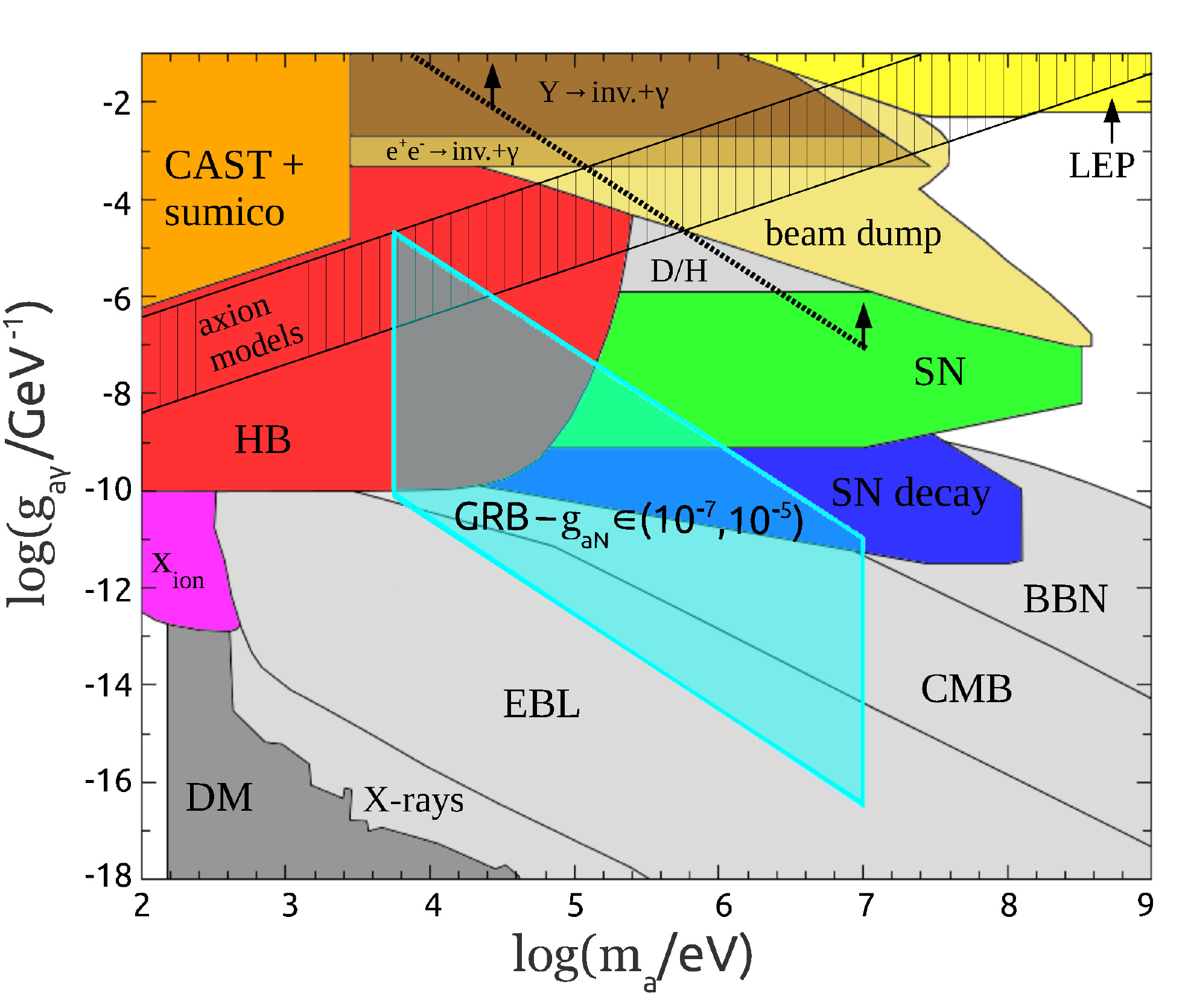}
\caption{\label{fig.exclu} Excluded regions in the $m_a-g_{a\gamma}$ space adapted and combined from previous studies \cite{hewett2012,pdg2016,cadamuro2012,jaeckel2017}, including the region explored in this work {based on} heavy axion decays to photons outside the GRB. The latter region, marked with {in a} transparent cyan color and labeled as ``GRB - $g_{aN}\in (10^{-7},10^{-5})$" can be ruled out for $g_{aN}$ within the {range of} values indicated. The brown and light brown regions have been excluded by collider and beam dump experiments \cite{beamdump1988,bauer2017}. The orange region is excluded {based on} searches of axion--photon conversions of solar axions with the ``CAST + Sumico" helioscope. The magenta region (``$x_{\rm ion}$") is ruled out since heavy axion decays would lead to a too early reionization of the universe \citep{arias2012}, {while} the dark gray region (``DM") is excluded because heavy axion dark matter would be excessively produced, as well as the light gray region labeled ``X-rays" because heavy axion decays inside galaxies would imply unobserved X-ray features \citep{cadamuro2012}. The region marked by black vertical lines corresponds to values of the standard ``axion models" and they are shown for illustration. {The dotted black line indicates the minimum values of $g_{a\gamma}$ required for heavy axions to power GRBs.} The remaining labels are described in Section \ref{sec:discu}.}
\end{figure*}

\begin{figure*}[tbp] 
\centering 
		\includegraphics[width=.9\textwidth,trim=0 5 0 0,clip]{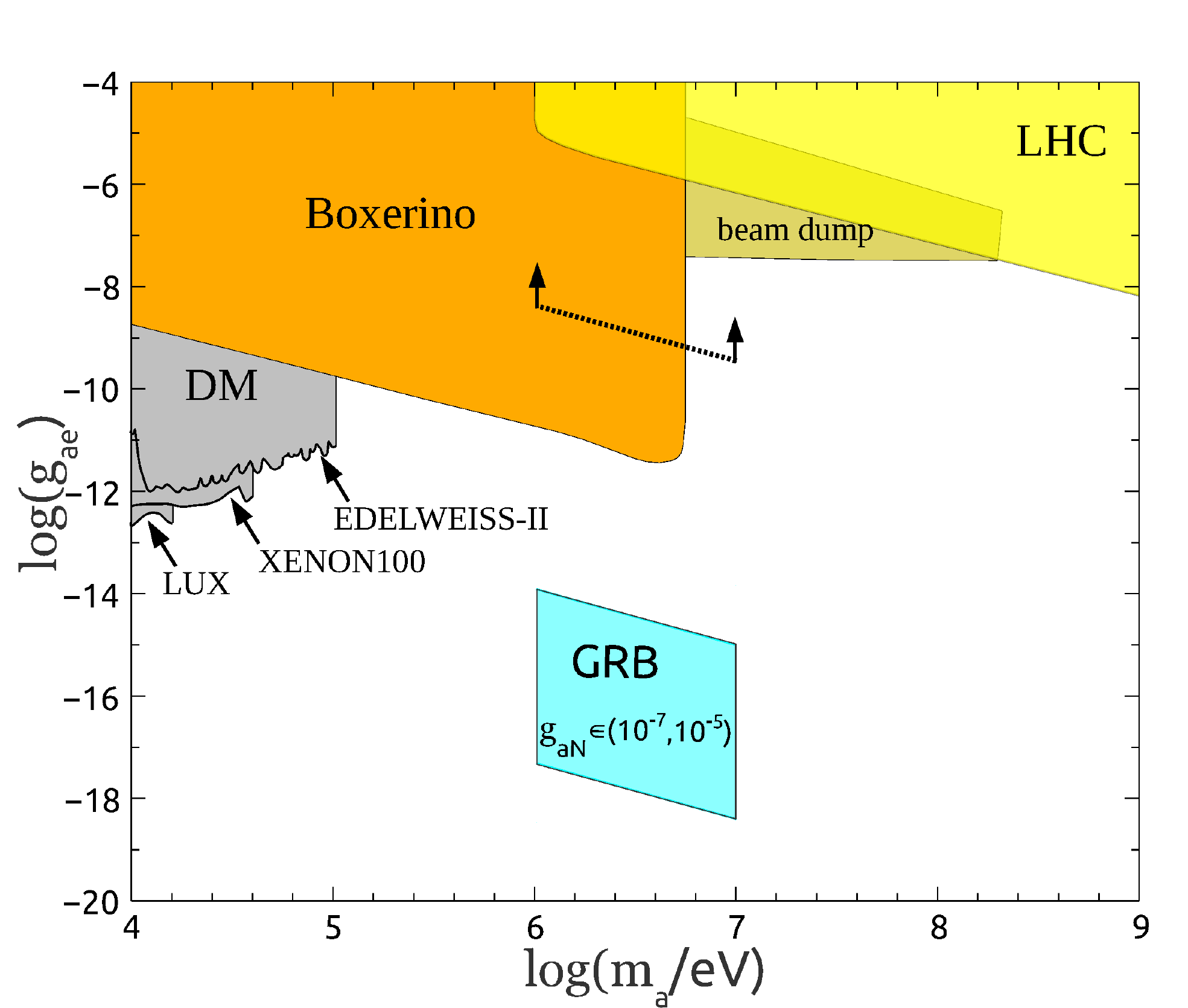}
\caption{\label{fig.exclu-e} Excluded regions in the $m_a-g_{ae}$ space adapted and combined from {previous studies} \cite{edel2013,xenon2017,lux2017,bauer2017}, including the region explored in this work {based on} heavy axion decays to electron--positrion pairs outside the GRB. The latter region, marked in cyan color and {labeled as} ``GRB - $g_{aN}\in (10^{-7},10^{-5})$", can be ruled out for $g_{aN}$ within the {range of} values indicated. The orange region {has been} excluded by searches {using} the Boxerino detector \cite{boxerino2012}, {while} the light brown region is excluded by beam dump experiments \cite{beamdump1988}. If galactic dark matter is {composed enterely of} heavy axions, {then} the gray region is excluded with the upper bounds given by the experiments indicated. We also include the region in transparent yellow, which is to be probed at LHC Run-2 searching for possible signals of  $Z\rightarrow a \, \gamma\rightarrow  e^+e^- $ decays \citep{bauer2017}. {The dotted black line indicates the minimum values of $g_{ae}$ required for heavy axions to power GRBs.}}
\end{figure*}


\section{Discussion}\label{sec:discu}

{In this study, we investigated} the structure of accretion disks in GRB {by considering} the cooling term that would arise due to heavy axion production via the nucleon--nucleon bremsstrahlung process. We found values of the coupling constant to nucleons for which the axions produced can escape from the disk by comparing the mean free path with the disk thickness. For instance, for heavy axions {with a mass} $m_a\sim 1 \, {\rm MeV}$, their coupling to nucleons can be $g_{aN}<10^{-5}$ and they could leave the disk without interacting for accretion rates $\dot{M}\lesssim 0.1 \, M_\odot{\rm s^{-1}}$, whereas for higher accretion rates  ($\dot{M}\lesssim 1 - 3 \, M_\odot{\rm s^{-1}}$), the disk is {denser} and it is necessary that $g_{aN}\lesssim 10^{-6}$ in order to have free streaming. In these cases, the structure of the disk {does not} depart significantly from the result corresponding to no axion production, although the slight changes are more noticeable for the higher values of $g_{aN}$ considered. 

The luminosity in heavy axions {can still be important and {it would lead to} a more efficient production of photons and/or $e^+e^-$ than {via} neutrino--antineutrino annihilation. This can be seen in Table \ref{table.lum}, especially for the highest values {considered for} $g_{aN}$.} Hence, as proposed {by} \cite{viaaxion2000} but having performed a more realistic calculation of the axion luminosity, if the {axions produced decay} to photons and/or $e^+e^-$ at distances $d_a<10^{9}{\rm cm}$, then the initial fireball could be generated and give rise to the observed GRB. {For example, this would be the case if} $g_{a\gamma}\gtrsim 10^{-5.1} \,{\rm GeV}^{-1}$ for dominant decays to photons and $g_{ae}\gtrsim 10^{-8.8}$ for dominant decays to $e^+e^-$, {with} $m_a\sim 2m_e$. {However, the former possibility {has been} excluded by beam dump and collider experiments \cite{beamdump1988,bauer2017}, constraints {based on} the primordial D/H ratio \citep{millea2015}, and the duration of SN 1987A. {Figure \ref{fig.exclu} shows} these and the remaining current bounds in the $g_{a\gamma}-m_a$ plane, {which are} valid {in} the case of heavy axions decaying to photons as the dominant channel. In this figure, {a black dotted line denotes} the required values of $g_{a\gamma}$ so that the typical decay distance of the heavy axions produced is $d_{a\rightarrow \gamma\gamma}= 10^{9}{\rm cm}$, {which shows} {that higher coupling values which would yield $d_{a\rightarrow \gamma\gamma}< 10^{9}{\rm cm}$, but they have already been excluded} for the relevant range of heavy axion mass.}

%
{On the other hand, the possibility of dominant decays to $e^+e^-$ to power GRBs {has not yet been excluded completely}. {This can be seen in Fig. \ref{fig.exclu-e}, where the Boxerino bound extends up to $m_a\simeq 5.6 \, {\rm MeV}$, and the black dotted line {denotes} the minimum values of $g_{ae}$ {required for} decays at distances shorter than $10^{9}\,{\rm cm}$, {thereby allowing} the possibility that heavier axions decay to $e^+e⁻$ {to create} the initial GRB fireball {provided that} $g_{ae}\lesssim 10^{-7.5}$, which has been excluded by beam dump experiments. We leave for future work a detailed study of any possible effects regarding axion production within the expanding fireball for these particular cases of $g_{ae}$ and $m_a$.} 


We also studied the cases in which heavy axions decay at longer distances. In the case that the decay to photons is dominant, the flux generated would lead to a clearly different energy dependence {compared with that typically observed, which is usually fitted by} the so-called Band model (Fig. \ref{fig.fluxes}). This would {occur} for $g_{aN}$ {taking} the values mentioned above and for the couplings $g_{a\gamma}$ and $g_{ae}$ within the shaded regions in Fig. \ref{fig.shaded} {(top panels)}. {In addition, we considered the possibility that} the decay channel to electron--positron pairs is dominant. In these cases, a flux of gamma rays would be generated by $e^\pm$ IC scattering on the CMB, and these photons would arrive {at the} Earth from a different direction {compared with that of the} usual GRB prompt emission{, which is explained by the deflection of} electrons and positrons in the intergalactic magnetic field. 

In Fig. \ref{fig.fluxes}, the GRB reference flux corresponds to a luminosity of $L_\gamma=10^{52}{\rm erg \ s^{-1}}$ in the energy band $(1\,{\rm keV}-40\,{\rm MeV})$ for a GRB at a redshift $z=1$. In fact, similar plots would be obtained for higher redshifts, and it {is} reasonable to expect that accretion disks with higher values of $\dot{M}$ become more feasible, {so} the conditions for heavy axion production would {remain at a} significant level even for $g_{aN}\sim 10^{-7}$.

{The lack of observations of the signatures described along with} the assumption that GRBs actually involve accretion disks with the physical conditions discussed \cite{pwf1999, kohri2002,kohri2005,chen2007,zalamea2011,janiuk2010,liu2015}, imply that either $g_{aN}\ll 10^{-7}$ and {there are} no restrictions on $g_{a\gamma}$ and $g_{ae}$, or if $10^{-7}\lesssim g_{aN} \lesssim 10^{-5}$, then the values of $g_{a\gamma}$ and $g_{ae}$ in the shaded regions of {the} plots in Fig. \ref{fig.shaded} should be excluded for $m_a=2m_e$ and $m_a=10\,{\rm MeV}$, {as well as} similar regions for intermediate values of $m_a$.
We note that {because} the IC emission generated by the $e^\pm$ would not be superimposed directionally and temporally to a normal GRB one, {then it can} be expected that even for lower levels of heavy axion production (i.e., $g_{aN}<10^{-7}$), {these} IC fluxes  would have been observed for the {aforementioned}  $g_{a\gamma}$ and $g_{ae}$ values. 
 
{In particular, {applying} these arguments to models with negligible or absent decays to $e^+e^-$ pairs and $m_a \sim 0.01-10 $ MeV, we can exclude the region of the $g_{a\gamma}-m_a$ space {marked in cyan color and labeled as} ``GRB - $g_{aN}\in (10^{-7},10^{-5})$" in Fig. \ref{fig.exclu}. {This region overlaps with existing bounds derived from different astrophysical arguments{, i.e.,} excessive energy loss in red giant stars would affect the observed counts in the horizontal branch of color--magnitude diagrams of globular clusters (`` HB"), the duration of the neutrino signal of {SN 1987A} (``SN"), and an unobserved delayed photon burst due to axion decay if they were produced in {SN 1987A} (``SN decay'')}. {Our region also overlaps with part of excluded regions by cosmological considerations, {i.e.}, heavy axion decays would have caused distortions in the CMB spectrum when the universe was opaque to photons (``CMB"), and the observed EBL flux cannot be exceeded (``EBL") by axion decay to photons when the universe became transparent. {In addition}, the decays to photons in the early universe cannot cause dilution of the neutrino density or affect the abundances of primordial nuclei that are consistent with Big Bang nucleosynthesis (``BBN"). {These constraints {derived from} cosmology, as discussed {by} \citep{jaeckel2017}, are model dependent in the sense that they are valid {provided that} the reheating temperature is relatively high ($T_{\rm RH}\gtrsim 120$ GeV) {in order to allow} significant thermal production of heavy axions. Since there are cosmological models {which involve} lower reheating temperatures, the bounds {based on} the cosmology arguments mentioned {above} (marked {in} light gray in Fig. \ref{fig.exclu}) would not apply, and then the results of the present work become more useful.} }

{In the case of dominant decays to $e^+e^-$, the existing bounds in the $m_a-g_{ae}$ space are shown in Fig. \ref{fig.exclu-e}. It can be seen {from this plot} that the region studied here, marked with cyan color and labeled ``GRB-$g_{aN}\in (10^{-7},10^{-5})$", has not been ruled out previously. We find that for $g_{aN}\in (10^{-7},10^{-5})$, this region should be excluded {because} {it implies} unobserved gamma ray flashes with an energy dependence {that is} clearly different from {that} corresponding to GRBs (i.e., the red lines in Fig.  \ref{fig.fluxes}).}

{To summarize our conclusions, we have seen that the production of heavy axions in GRB accretion disks can lead to the formation of the GRB fireball {via} the decay channel $a\rightarrow e^+e^-$, {because an allowed region in the $m_a - g_{ae}$ space is} still compatible with this situation (Fig. \ref{fig.exclu-e}). {By constrast, heavy axions decaying to photons cannot} be the origin of the fireball, {because} the required values {for} $g_{a\gamma}$ {have been} excluded (Fig. \ref{fig.exclu}). Finally, we have found other combinations of values for the couplings $g_{aN}$, $g_{a\gamma}$, and $g_{ae}$, which as discussed above, {are} inconsistent with the current understanding and observations of GRBs. This can be used as %
a complementary tool to constrain models that involve heavy axion-like particles.}

\section*{Acknowledgments}
I thank O.A. Sampayo, G.E. Romero, and G.S. Vila for helpful comments.
I also thank CONICET (PIP-2013-2015 GI 160) and Universidad Nacional de Mar del Plata for financial support.  



\end{document}